\documentclass[reprint,prb,showpacs]{revtex4-1}
\usepackage{bm}
\usepackage{graphicx}
\usepackage[usenames]{color}
\usepackage{amsmath,amsfonts,amssymb}
\usepackage{hyperref}

\begin{document}
\title{Topological Order Parameters of the Spin-1/2 Dimerized Heisenberg Ladder
in Magnetic Field}

\date{\today}

\author{Toshikaze Kariyado}\email{kariyado@rhodia.ph.tsukuba.ac.jp}
\author{Yasuhiro Hatsugai}\email{hatsugai@rhodia.ph.tsukuba.ac.jp}
\affiliation{Division of Physics, Faculty of Pure and Applied Sciences,
University of Tsukuba, Tsukuba, Ibaraki 305-8571, Japan}


\begin{abstract}
 Topological properties of the spin-1/2 dimerized Heisenberg ladder are
 investigated focusing on the plateau phase in the magnetic
 field whose magnetization is half of the saturation
 value. 
 Although the applied magnetic field removes most of the
 symmetries of the system, there is a symmetry protected
 topological phase supported by the spatial inversion symmetry. The
 $Z_2$ Berry phase associated
 with a symmetry respecting boundary and quantized into $0$
 and $\pi$ is used as a
 symmetry protected topological order parameter. Edge states are also
 analyzed to confirm the bulk--edge correspondence. In addition, a
 symmetry breaking boundary is considered. Then, we observe a unique
 type of quantization of
 the Berry phase, a quantization into $\pm\pi/2$ of the Berry phase. In
 this case, the bulk--edge correspondence is also unique, namely, there
 emerge ``polarized'' edge states for the case with $\pm\pi/2$
 quantization. We also evaluate the entanglement entropy by the infinite
 time-evolving block decimation (iTEBD) to complement the Berry phase
 based arguments. Further, a different type of the
 topological order parameter is extracted from the matrix product state
 representation of the ground state given by the iTEBD.
\end{abstract}

\maketitle

\section{Introduction}
Recent use of topology in condensed matter substantially revised our view on the
characterization of materials especially for the gapped
phases\cite{Wen89,Hatsugai05-char,doi:10.1143/JPSJ.75.123601}.
One of the advantages of a topological view is robustness
against continuous deformation. One can use a topological quantity
to distinguish topological phases since the quantized nature of it
guarantees its invariance against continuous deformation
\cite{Thouless98book,doi:10.1143/JPSJ.75.123601,PhysRevB.82.155138}.
Due to the theorem by von Neumann-Wigner, however, 
truly generic phase can be a single class since any generic
states can be adiabatically connected.
Then, the symmetry restriction is essential.
The symmetry can be gauge symmetry,
time-reversal, particle-hole,
reflection, etc\cite{doi:10.1143/JPSJ.75.123601,
  PhysRevB.78.195125,PhysRevB.82.155138,PhysRevB.85.075125}. 
When these restrictions give rise to a new nontrivial
  phase, it is a symmetry protected topological (SPT) phase.
A typical SPT order parameter is the Berry phase, which
takes any value without symmetry, but can be quantized with the
appropriate symmetry.

The continuous deformation, or adiabatic continuation, is also essential for
establishing the bulk--edge correspondence\cite{Hatsugai93b,Ryu02,HatsuSSC,PhysRevB.90.085132},
which is one of the fundamental concepts for characterizing topological
phases.
Physics at the bulk and the edges are not independent and related
each other especially for the gapped case. 
Introduction of a boundary sometimes
breaks symmetries
of the bulk system, but sometimes does not. From the viewpoint of the
symmetry protection, whether the edge respects a bulk
symmetry or not has a special importance. 

In this paper, topological properties of a dimerized spin-1/2
Heisenberg ladder with antiferromagnetic coupling is investigated
focusing on the plateau phase at half of the saturation,
which we call a 1/2-plateau phase. The 1/2-plateau phase appears in the
applied magnetic field, which breaks most of the symmetry of the
system. The ladder model itself has been studied
extensively\cite{PhysRevB.47.3196,PhysRevLett.79.5126,refId0,PhysRevLett.82.1768,PhysRevB.62.14965,PhysRevB.77.224433,PhysRevB.83.104423},
and some of the studies shed lights on the topological
aspects of the ladder model\cite{PhysRevB.79.115107,PhysRevB.79.205107,PhysRevB.86.195122}, but the focus was mainly on the case
without magnetic field. (Very recently, plateau phases at finite
magnetization in spin chains are studied by a SPT viewpoint in Ref.~\onlinecite{chainSPT}.) A main purpose of this paper is
to show the 1/2-plateau phase here is a SPT phase protected by the spatial inversion symmetry
that survives even with the finite external magnetic field. The Berry phase and
the entanglement entropy are employed to characterize a SPT phase. As we
will explain later,
boundary shapes are essential for both of the Berry phase and the
entanglement entropy, and the boundary that keeps the inversion
symmetry is mainly used not to destroy the symmetry
effects. 
Importance of the symmetry is also demonstrated by introducing artificial
symmetry breaking, and by a spontaneous symmetry breaking caused by a
ring exchange. We further study edge states to establish the bulk--edge
correspondence. In order to complement the above arguments, a boundary
that breaks inversion symmetry is also treated. Naively, one may think
that such a symmetry
breaking boundary is not useful for characterizing symmetry protected
topological phases. However, for a specific type of symmetry
breaking boundary, we found a unique quantization of the Berry phase,
i.e., a fractional quantization of the Berry phase into $\pm\pi/2$,
instead of the widely observed 0/$\pi$-quantization. Further, it is shown
that edge states are also unique for the $\pm\pi/2$-quantization case:
there appear ``polarized'' edge state such that up and down
spins are localized at the opposite ends of the finite system. 
The applied magnetic field is essential for achieving the fractional
quantization.

This paper is organized as follows. In Sec.~\ref{sec:model_methods}, a
model Hamiltonian and physical quantities used in our
topological characterization are introduced. Then, the numerical methods
to obtain those physical quantities, the exact diagonalization and
the infinite time-evolving block decimation\cite{PhysRevLett.98.070201}
(iTEBD), are
explained. There, we also explain how
the topological character of the
system is encoded in the matrix product state, in terms of the
transformation law against the symmetry operation.
Section~\ref{sec:results_discussions} contains main results
of this paper. First, the magnetization curve is shown to take a glance
at the plateau phase on which we focus in this paper. After that, the
topological properties of the 1/2-plateau phase, such as the quantized
Berry phase and the bulk-edge correspondence are
discussed in detail. The effects of ring exchange is also
considered. Finally, we make a comparison between the 0-plateau phase
and the 1/2-plateau phase. The paper is summarized in
Sec.~\ref{sec:summary}. 

\section{Model and Methods}\label{sec:model_methods}
\begin{figure}[tbp]
 \begin{center}
  \includegraphics[scale=1.0]{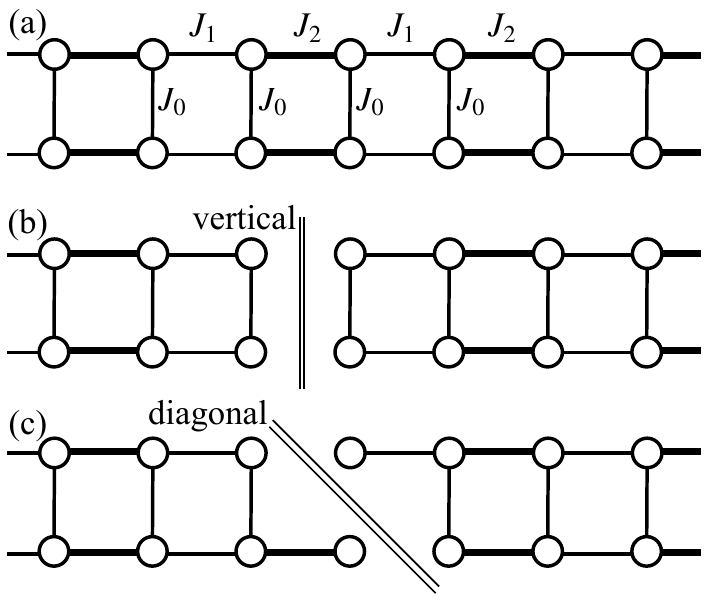}
  \caption{(a) The dimerized spin-1/2 Heisenberg ladder. Definitions of
  parameters are shown. (b) The vertical edge. (b) The diagonal edge.}
  \label{fig:model}
 \end{center}
\end{figure}

The model treated in this paper is a dimerized spin-1/2 Heisenberg
ladder\cite{PhysRevLett.82.1768} with the Zeeman field, whose
Hamiltonian is written as 
\begin{multline}
  \hat{H}=
  \sum_{i=1}^{L}\sum_{j=1,2}
 \Bigl[
 J_1\bm{S}_{2i,j}\cdot\bm{S}_{2i+1,j}
 +J_2\bm{S}_{2i+1,j}\cdot\bm{S}_{2i+2,j}
 \Bigr]\\
 +J_0\sum_{i=1}^{2L}\bm{S}_{i,1}\cdot\bm{S}_{i,2}
 -B_z\sum_{i=1}^{2L}\sum_{j=1,2}S^{z}_{i,j}. \label{eq:Hamiltonian}
\end{multline}
(See Fig.~\ref{fig:model}.) Here, we concentrate on
the antiferromagnetic coupling, namely all of $J$s in
Eq.~\eqref{eq:Hamiltonian} are assumed to be positive. For convenience,
a parameter $\Delta$ is introduced as $\Delta=(J_2-J_1)/2$. If
$\Delta\neq 0$, the minimum unit cell is composed of four spins, while if
$\Delta=0$, it is composed of two. This unit cell structure is
essential to obtain the 1/2-plateau phase\cite{PhysRevLett.78.1984} that
we concentrate on. When $B_z=0$, the system has the rotational symmetry in the
spin space and the time reversal symmetry, but those symmetries are
broken for finite $B_z$.
However, even with finite $B_z$, the system retains the
spatial inversion symmetry, which is essential for protecting the 
topological phase in the 1/2-plateau phase. 

In order to elucidate the topological properties of the model, we
calculate the magnetization, the Berry phase, and the entanglement
entropy. The magnetization is calculated to show the existence
of the plateau phase. The Berry phase
defined below works as a
symmetry protected topological order parameter to identify two
topologically distinct states having the same
symmetry\cite{doi:10.1143/JPSJ.75.123601}. 
The entanglement entropy for the
spatial bipartition is evaluated to check whether the two states can be
smoothly connected or not. Also the entanglement entropy gives
a picture of the bulk-edge correspondence because the topological
character
of the system is encoded in the entanglement
spectrum\cite{PhysRevB.81.064439,PhysRevB.84.205136}, and the
entanglement entropy contains a contribution from the edge
states\cite{PhysRevB.73.245115,PhysRevB.90.085132}.
Analysis based on
the transformation law of the matrices in the matrix product state (MPS)
representation is also performed to complement the Berry phase
based arguments, i.e., we extract a topological order parameter other
than the quantized Berry phase from the MPS representation. In addition, by
making use of
the translationally invariant
MPS representation that enables us to access the thermodynamic limit, we
can discuss the collapse of the SPT phase with spontaneous symmetry
breaking. 

\subsection{Numerical methods}
The calculations of the Berry phase and the investigation of the edge
states are performed by the exact diagonalization of the finite size
system. The ground state energy and the wave function are numerically
evaluated by the Lanczos algorithm combined with the inverse iteration
method. In order to define the Berry phase, we apply a
local gauge twist on the bonds (possibly on the multiple bonds) at the
given boundary as 
\begin{equation}
 S^+_{L}S^-_{R}+\text{h.c.} \longrightarrow \mathrm{e}^{\mathrm{i}\phi}S^+_LS^-_R+\text{h.c.},
\end{equation}
where $L$ and $R$ denote the sites on the left and right sides of the
given boundary. Using the ground state wave function $|G_\phi\rangle$
at each $\phi$, the Berry phase $\gamma$ is obtained as
\begin{equation}
 \mathrm{i}\gamma=\int_0^{2\pi}\mathrm{d}\phi\langle
  G_\phi|\partial_\phi|G_\phi\rangle.
\end{equation}
 In practice, by discetizing the range [0,$2\pi$] as
 $\phi_i=2\pi{i}/N$ ($i=1,\dots,N-1$), it is evaluated as \cite{doi:10.1143/JPSJ.75.123601,doi:10.1143/JPSJ.74.1674,PhysRevB.78.054431}
\begin{equation}
 \gamma = \textrm{arg}\biggl(\langle G_{\phi_{N-1}}|G_{\phi_0}\rangle\prod_{i=0}^{N-1}\langle G_{\phi_i}|G_{\phi_{i+1}}\rangle\biggr).
\end{equation}
Because the gauge twist is applied on the bonds crossing the boundary, the Berry phase
depends on the boundary shape, which is essential for establishing the
bulk-edge
correspondence\cite{PhysRevB.88.245126,PhysRevB.90.085132}. With the
exact diagonalization scheme, finite
size effects are unavoidable. However, as far as the Berry phase is
quantized due to some symmetry, as we will see soon later, there is
practically no size effect on the numerically obtained Berry phases.

In order to obtain the entanglement entropy and the MPS,
the iTEBD
method is employed\cite{PhysRevLett.98.070201}. In the iTEBD, a
translationally invariant MPS representation of the ground state is
iteratively obtained. It becomes exact in the large $\chi$ limit,
where $\chi$ is truncation dimension corresponding to the dimension of
the matrix in the MPS representation. For gapped phases, small $\chi$ is
sufficient to
obtain results in practical precision. An advantage of the iTEBD is
accessibility to the thermodynamic
limit, i.e., it is free from the finite size effect, because it
provides a translationally invariant MPS representation by
construction. The finite size effect free nature is essential for
discussing spontaneous symmetry breaking. To
perform the iTEBD, translation symmetry is essential, and the target
system is assumed to be composed of repetition of unit objects. Then,
the iTEBD naturally gives entanglement entropy for
the bipartition such that the system is divided into two parts in
between the two neighboring unit objects. In the following, unit
objects are appropriately chosen so as to make a given boundary in
between two of them. In this way, the entanglement
entropy depends on shapes of given boundary as it should do.

With the iTEBD, it is possible to obtain the translationally invariant
canonical MPS representation\cite{PhysRevB.78.155117} of the state such
that 
\begin{equation}
 |\Psi\rangle=\sum_{\{s_i\}}\cdots
  \Gamma_{s_i}\Lambda\Gamma_{s_{i+1}}\Lambda\cdots
  |\cdots,s_i,s_{i+1},\cdots\rangle,
\end{equation}
where $s_i$ denotes the labels of local states, and $\Gamma_s$ and
$\Lambda$ are $\chi\times\chi$ matrices. $\Lambda$ is a diagonal matrix
whose entries are nonnegative, and related to the
entanglement entropy $S$ as 
\begin{equation}
 S=-\sum_i\lambda_i^2\log\lambda_i^2,
\end{equation}
where $\lambda_i$ is the diagonal elements of $\Lambda$.
When $|\Psi\rangle$ respects some
symmetry, $\Gamma_s$ must react against the symmetry operation
appropriately. Then, physical states with the same symmetry can be
classified by this transformation law of $\Gamma_s$. When
the symmetry operation is written by a product of local operators
$O^{(a)}$, $\Gamma_{s'}$ transforms as\cite{PhysRevLett.100.167202}
\begin{equation}
 \sum_{s'}O^{(a)}_{ss'}\Gamma_{s'}
  =\mathrm{e}^{\mathrm{i}\theta_a}U^\dagger_a\Gamma_sU_a,
  \label{eq:trans}
\end{equation}
where $U_a$ is an unitary matrix that satisfies
$[\Lambda,U_a]=0$. Mathematically, $U_a$ gives a projective
representation of the symmetry
operation, and the states are distinguished by its factor
set\cite{PhysRevB.81.064439,PhysRevB.86.125441,PhysRevB.90.235111}. If the
operation involves the spatial inversion
symmetry, which plays a central role in this paper, $\Gamma_{s'}$ at the
left hand side of Eq.~\eqref{eq:trans} is replaced by ${}^t\Gamma_{s'}$
that is the transpose of $\Gamma_{s'}$\cite{PhysRevB.81.064439}, since the inversion operation
reverses the order of $s_i$. 
In general, a cyclic group
generated by a single element, like the case that there is only
inversion symmetry, leads no interesting factor set. However, this
transposition makes the inversion symmetry useful in classification of
the states. Namely, there is a restriction on $U_a$ for the spatial
inversion symmetry (denoted as $U_I$ hereafter) such that 
${}^{t}U_I=\pm U_I$, which means that $U_I$ should be symmetric or
antisymmetric\cite{PhysRevB.81.064439}. For antisymmetric $U_I$, the
relation $[\Lambda,U_I]=0$ gives degeneracy of the entanglement
spectrum, in which the topological
properties of the system is
encoded\cite{PhysRevB.81.064439}. Particularly, if the
entanglement spectrum is at least doubly degenerate, the entanglement
entropy has a lower bound $\log 2$.
As we will focus on the phase with the finite magnetic field, in
which the time reversal symmetry and the symmetries in the spin space
are not effective, $\zeta$, which is defined according to 
\begin{equation}
 {}^{t}U_I=\zeta U_I,\label{eq:def_zeta}
\end{equation}
 and takes values of $+1$ and $-1$, is employed as a topological order
 parameter to classify the
 phases. 

In practice, $U_I$ is obtained as an ``eigenmatrix'' of a linear matrix
map\cite{PhysRevLett.100.167202} 
\begin{equation}
 \mathcal{E}_I(U)=\sum_{ss'}\hat{O}^{(I)}_{ss'}({}^t\Gamma_s\Lambda)U(\Gamma_{s'}\Lambda)^\dagger,
\end{equation}
whose eigenvalue $\epsilon$ satisfies $|\epsilon|=1$. In specific, $U_I$
satisfies
\begin{equation}
 \mathcal{E}_I(U_I)=\mathrm{e}^{\mathrm{i}\theta}U_I.
\end{equation}
Numerically, $\mathrm{e}^{\mathrm{i}\theta}$ and $U_I$ are obtained as
an eigenvalue and an eigenvector of the matrix representation of the
map $\mathcal{E}_I$, whose matrix elements are defined as
\begin{equation}
 T_{ij;i'j'}
  =\sum_{ss'}\hat{O}^{(I)}_{ss'}({}^t\Gamma_s)_{ii'}\lambda_{i'}(\Gamma_{s'})_{jj'}^*\lambda_{j'}.
\end{equation}
The matrix $T$ works as a transfer matrix
when we calculate the overlap between the wave functions before and
after the symmetry operation is applied\cite{PhysRevB.86.125441}. It
means that, as far as the state respects the symmetry, the largest norm
of the eigenvalues of $T$ becomes unity. Then $U_I$ is obtained from an
eigenvector associated with the eigenvalue with unit norm, and $\zeta$
is reduced from it. On the other hand, if the largest norm of the
eigenvalues of $T$ is less than unity, it implies that the state under consideration
does not respect the symmetry. In other words, $T$ has an ability to
detect whether a given state is invariant against a symmetry
operation. For convenience, we set $\zeta$ to zero when $T$ detects a
symmetry breaking and $U_I$ is unavailable.

\section{Results and Discussions}\label{sec:results_discussions}
\begin{figure}[tbp]
 \begin{center}
  \includegraphics[scale=1.0]{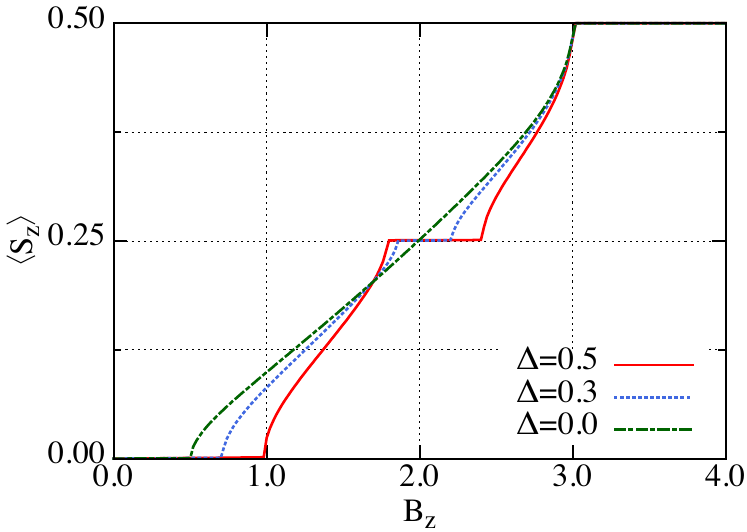}
  \caption{(Color online) The magnetization curve for $\Delta=0.0$,
  $0.3$, and $0.5$
  obtained with $\chi=24$. Magnetization is already converged with
  $\chi=24$ or smaller.}\label{fig:magnetization}
 \end{center}
\end{figure}
The magnetization curve obtained with the iTEBD for $J_0=1.0$,
$J_1=1.0-\Delta$, and $J_2=1.0+\Delta$ with several values of $\Delta$
is shown in Fig.~\ref{fig:magnetization}. The essential features of the
curves are consistent with those in Ref.~\onlinecite{PhysRevLett.82.1768}.
Namely, there are two plateau phases, for $\langle S_z\rangle=0$ (0-plateau) and 
$\langle S_z\rangle=1/4$ (1/2-plateau). The width of the 1/2-plateau
depends on the
strength of the dimerization $\Delta$. If there is no dimerization, the
unit cell contains only two sublattices, and the plateau at $\langle
S_z\rangle=1/4$, which is a half of the saturation magnetization, is not
allowed. On the other hand, the
0-plateau phase does not vanish in the zero dimerization limit. This is
natural because the 0-plateau phase
without dimerization is actually in the rung singlet
phase\cite{PhysRevB.67.100409,PhysRevB.74.155119}, and the rung
singlet is expected to be stable for small dimerization.

\subsection{Topological order parameter in the 1/2-plateau phase}
\begin{figure}[tbp]
 \begin{center}
  \includegraphics[scale=1.0]{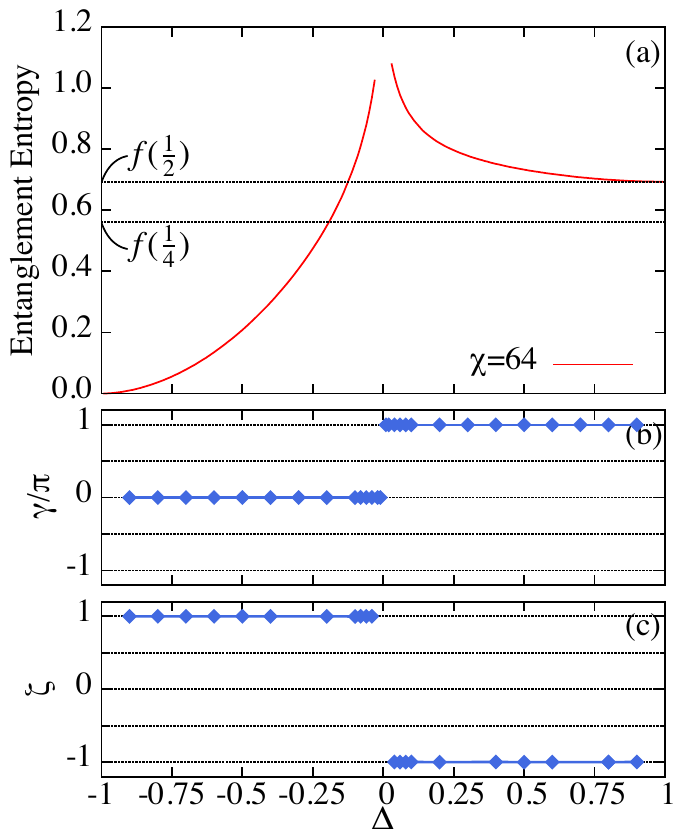}
  \caption{(Color online) (a) The entanglement entropy, (b) the Berry
  phase, and (c) the other topological order parameter $\zeta$ as a
  function of $\Delta$ for the vertical edge. The entanglement entropy is calculated with $B_z=2.0$ and $\chi=64$. The Berry
  phase is obtained with the $\langle S_z\rangle=1/4$
  sector in the six unit cell (24 spins) system. $\zeta$ is
  calculated with $\chi=32$.}\label{fig:type1}
 \end{center}
\end{figure}

Let us move on to
the topological properties of the 1/2-plateau
phase. Figures.~\ref{fig:type1}(a) and
\ref{fig:type1}(b) show the numerically obtained Berry phase and entanglement entropy
for the 1/2-plateau phase. These quantities depend on the boundary shape, and
the vertical edge in
Fig.~\ref{fig:model}(b) is employed here. (Later we will discuss the edge
states and then the diagonal edge is also used.) What is important
is that the vertical edge does not break the inversion symmetry whose
inversion center is at the boundary. Then, the Berry phase is
quantized into 0 or $\pi$ as in
Fig.~\ref{fig:type1}(b). This quantization is caused by the spatial
inversion symmetry\cite{PhysRevB.88.245126}. Here, the inversion
symmetry means that the
Hamiltonian with the gauge twist $\phi$ satisfies 
\begin{equation}
 \hat{H}_{-\phi}=\hat{P}^{-1}\hat{H}_\phi\hat{P}
\end{equation}
where $\hat{P}$ is an appropriate unitary matrix. This relation combined
with the assumption that the ground state is unique implies $\gamma=-\gamma$ (mod $2\pi$), which immediately leads to the
quantization of $\gamma$ into $0$ or $\pi$. As
this quantization is robust provided that the spatial inversion symmetry
is kept intact, $\gamma$ can be regarded as a symmetry protected
topological order parameter. The Berry phase $\gamma$ shows a jump at
$\Delta=0$, which suggests a phase transition. This phase
transition is also detected by the entanglement entropy as its divergent
behavior. 

Analysis of the transformation law of $\Gamma_s$ also supports this
conclusion. In fact, as
is shown in Fig.~\ref{fig:type1}(c), $\zeta$ is 1 for $\Delta<0$ and $-1$
for $\Delta>0$, which 
confirms that the both states respect the spatial
inversion symmetry, but are topologically distinct from each
other. For $\Delta>0$, the entanglement entropy is
always larger than $\log 2$, due to the double degeneracy of the
entanglement spectrum that is associated with $\zeta=-1$. Importantly, $\zeta=-1$
is found in the phase with $\gamma=\pi$, which implies consistency
between $\gamma$ and $\zeta$ as a topological order parameter. Since we
are considering the finite $B_z$ case, the time reversal symmetry and
the rotational symmetry in spin space cannot be used here.

For this classification of the phases, choice of the position of the 
boundary is essential because the states with $\Delta>0$ and that with
$\Delta<0$ are equivalent by a single site translation of the whole
system. This means that the observed transition is similar to that of
the Su-Schrieffer-Heeger model\cite{PhysRevLett.42.1698}, or of the
dimerized spin-1/2 Heisenberg
chain\cite{PhysRevB.45.2207,PhysRevB.71.054413,PhysRevB.87.054402,0953-8984-26-45-456001}. A relation to the dimerized chain is understood by considering
the strong rung coupling limit, though our calculation so far assumes
the comparable rung and leg couplings. The spin-1/2 ladder in the strong
rung coupling limit at $\langle S_z\rangle=1/4$ is effectively described by a XXZ chain model
at {\it zero} magnetization with easy plane
anisotropy\cite{refId0,PhysRevB.57.3454,0953-8984-13-33-321}. Then, the
present phase under consideration is expected to be connected to the
phase of the dimerized XXZ chain. On the other hand, the dimerized XXZ chain with easy plane
anisotropy is smoothly connected to the anisotropic chain. (This
is not the case with the easy axis, or Ising, anisotropy since the
system will be in an antiferromagnetic phase
at least for the weak dimerization limit.) Now, we
confirm that the
transition in Fig.~\ref{fig:type1} is the same type as the transition
in the dimerized spin-1/2 {\it chain} with
{\it no} external magnetic
field\cite{PhysRevB.87.054402,0953-8984-26-45-456001}, where two
distinct phases are distinguished by the positions of spin singlets.

\subsection{Role of the symmetry in protecting the topological phases}
\begin{figure}[tbp]
 \begin{center}
  \includegraphics[scale=1.0]{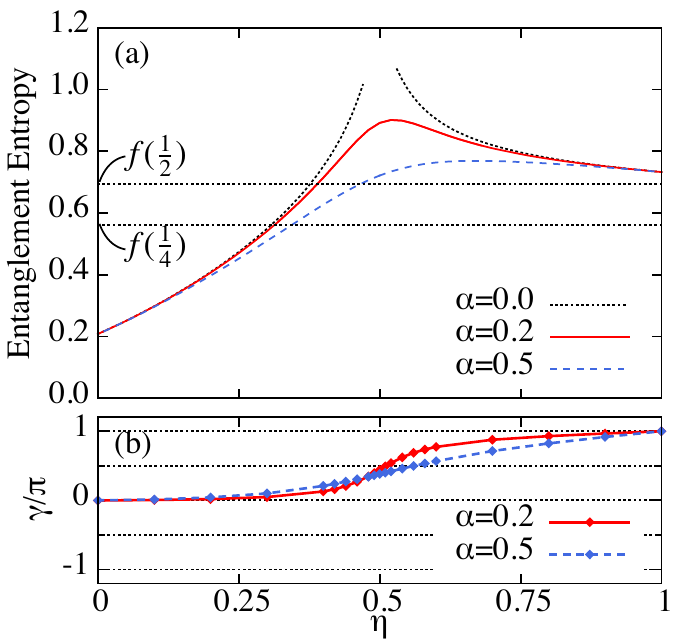}
  \caption{(Color online) (a) The entanglement entropy and (b) the Berry
  phase with
  artificial symmetry breaking. The entanglement entropy for
  $\alpha=0.2$ and $0.5$ is obtained with $B_z=2.0$ and
  $\chi=24$. The Berry phase is
  calculated with the $\langle S_z\rangle=1/4$ sector
  of six unit cell (24 spins) system.}
  \label{fig:artifitial}
 \end{center}
\end{figure}
In the previous section, the states are characterized by the
Z$_{\text{2}}$ Berry phase (and the
transformation law of $\Gamma_s$). 
In order to demonstrate the role of the symmetry, we introduce
symmetry breaking term and show the two phases are continuously
connected if the
symmetry is broken. Here, we add a term
\begin{equation}
\hat{H}_{\text{artificial}}=\delta J_0\sum_{i=1}^{2L}(-1)^i\bm{S}_{i,1}\cdot\bm{S}_{i,2},
\end{equation}
which makes staggered modulation of the rung
coupling\cite{0953-8984-18-40-014,Nemati:2011aa,Li20132422} and breaks
the spatial inversion symmetry whose inversion
center is in between the two rungs. Parameters $\eta$ and $\alpha$ are introduced as
\begin{align}
 \Delta&=\eta-0.5,\\
 \delta J_0&=\alpha\eta(1-\eta).
\end{align}
With this definition, the states with $\eta=0$ ($\Delta=-0.5$,
$\delta J_0=0$) and 
$\eta=1$ ($\Delta=0.5$, $\delta J_0=0$) retain the spatial inversion
symmetry. 
The Berry phase and the entanglement entropy for
several $\alpha$ in the 1/2-plateau phase are shown in
Fig.~\ref{fig:artifitial}. The Berry phase
is no longer quantized and the jump observed in Fig.~\ref{fig:type1} is
removed. (In principles, the
finite size effects come into play in this case
without quantization, but in the present case, the
results obtained with the 20 spin system and the 24 spin system are
nearly identical.) At the same time, divergence in the entanglement
entropy is also removed, similar to the case of the symmetry broken
dimerized chain\cite{0953-8984-26-45-456001}. Furthermore, $\zeta=0$ since the
maximum norm of the
eigenvalues of $T$ is less than unity as it should be with the broken
spatial inversion symmetry. All of these results are consistent with the
crucial role of the symmetry. Two kinds of symmetry
protected topological order parameters, $\gamma$ and $\zeta$, which pick
up the same information when the inversion symmetry exists, give quite
different information when the inversion symmetry is broken. By
definition, $\zeta$ is set to zero in such a case. On the other
hand, $\gamma$ takes some value and works as a measure of the
``distance'' to the topological phase or the trivial phase.

\subsection{Bulk--edge correspondence for a symmetry preserving boundary}
\begin{figure}[tbp]
 \begin{center}
  \includegraphics[scale=1.0]{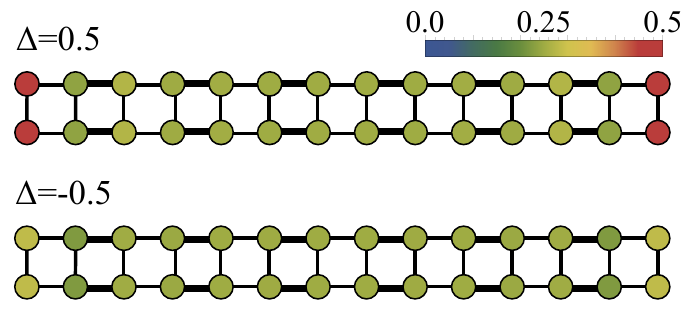}
  \caption{(Color online)
  Site resolved magnetization with the
  vertical open boundary for $\Delta=0.5$ (upper panel) and
  $\Delta=-0.5$ (lower panel).}\label{fig:edge1}
 \end{center}
\end{figure}
Next, we move on to the analysis of the edge states to see the bulk--edge
correspondence. In the following, we see that $\gamma=\pi$ ($\zeta=-1$)
state shows a clear sign of an
edge state, while $\gamma=0$ ($\zeta=1$) state does not. 
Figure~\ref{fig:edge1} shows
the site resolved magnetization of the ground state obtained with the
open boundary. For this calculation, we used $\Delta=\pm{0.5}$ and
$B_z=2.0$. For $\Delta=0.5$, the total magnetization of the ground state
deviates from exactly being $\langle S_z\rangle=1/4$, but the ground
state has one more extra up spin. This extra up spin is localized at the
boundary, as we can see from Fig.~\ref{fig:edge1}, which makes the edge
distinct from the bulk. No significant change of the local magnetization
is found in the bulk part. On the other hand, for $\Delta=-0.5$, the
ground state magnetization exactly satisfies $\langle S_z\rangle=1/4$,
and the local magnetization is only weakly affected at the
boundary. These behaviors are clearly explained in the $\Delta=\pm{1}$
limit, i.e., the decoupled limit. In this limit, we only have to
consider the four site clusters. For $\Delta=-1$, the boundary does not
break a cluster, and we expect no edge states. On the other hand, for
$\Delta=1$, a cluster at the boundary is broken, and the lowest energies
of the broken cluster at each magnetization are obtained as
$-2B_z+J_0/2$ ($\langle S_z\rangle=1/2$), $-B_z-J_0/2$ (
$\langle S_z\rangle=1/4$), and $-3J_0/2$ ($\langle S_z\rangle=0$). In the
present case, $J_0=1.0$ and $B_z=2$, the fully polarized state
($\langle{S_z}\rangle=1/2$) is chosen at the boundary, which indicates
that the extra up spin is localized at the boundary. The $\Delta=\pm{0.5}$
cases are adiabatically connected to the case with $\Delta=\pm{1}$.

Existence of the edge states is also reflected in the entanglement
entropy. This point is clarified by considering $\Delta=\pm 1$.
In this case, it is easy to evaluate the entanglement entropy because we
only have to
take account of the entanglement within a single four-site cluster. For
$\langle S_z\rangle=1/4$ and $\Delta=1$, the resultant entanglement
entropy is $f(1/2)$, where $f(x)=-x\log x-(1-x)\log(1-x)$. (See
Appendix~\ref{appendix}.) For $\Delta=1$, because of
$\zeta=-1$, the degeneracy of the entanglement spectrum is expected, and
$f(1/2)=\log 2$ is consistent with the lower bound set by the degeneracy.
On the other hand, for $\Delta=-1$, where no edge state is expected, 
the entanglement entropy is zero, since the cluster is not broken. 
In fact, the
numerical result in Fig.~\ref{fig:type1}(a) shows that the entanglement
entropy approaches to these values in $\Delta=\pm 1$ limit. In short, the
entanglement entropy in positive $\Delta$ side is largely contributed
from the edge state. 

\subsection{Symmetry breaking boundary and fractional quantization of the Berry phase}
\begin{figure}[tbp]
 \begin{center}
  \includegraphics[scale=1.0]{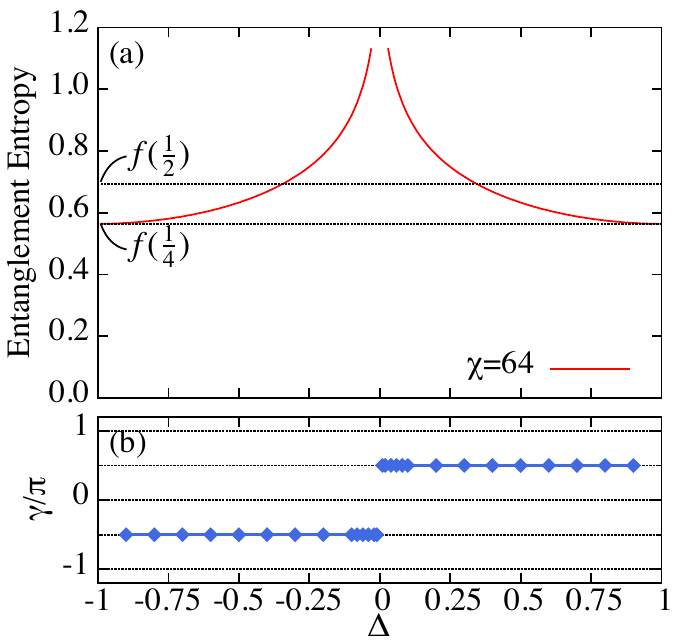}
  \caption{(Color online) (a) The entanglement entropy and (b) the Berry
  phase for the
  diagonal edge. The entanglement entropy is calculated
  with $B_z=2.0$ and $\chi=64$. The Berry phase is obtained using the
  system with six unit cells (24 spins) with $\langle
  S_z\rangle=1/4$.}\label{fig:type2}
 \end{center}
\end{figure}
So far, we have only considered the vertical edge. Now, let us
move on to the diagonal edge. [See Fig.~\ref{fig:model}(c)]. Important
feature of the diagonal edge is that it breaks the inversion symmetry no
matter where we choose as the inversion center even if
the bulk symmetry is kept intact. (Recall that the
vertical edge keeps the inversion symmetry if we set the inversion
center at the boundary.) 
Even
in this case, since the bulk symmetry is preserved and the Berry phase and
the entanglement entropy are bulk quantities, i.e., both quantities are
obtained by bulk ground state wave functions, these quantities still
have an ability to sense a topological phase transition.
But, of course, they should behave differently from the
case of the symmetry preserving boundary. For instance, the Berry
phase need not be quantized into 0 or $\pi$, and not
necessarily useful in a naive thought.
However, in the present model, the diagonal edge also exhibits
interesting phenomena as shown below. First, the Berry phase
for the diagonal edge in
the 1/2-plateau phase is shown in
Fig.~\ref{fig:type2}. There, we can see that the Berry phase is
quantized into $\pm\pi/2$, that is, the
Berry phase shows the unique fractional
quantization\cite{PhysRevB.90.085132}. This fractional
quantization obeys from the formula 
\begin{equation}
 \gamma_{\text{diagonal}}
  =\gamma_{\text{vertical}}-2\pi(S-\langle S_{z}\rangle),
  \label{gamma_trans}
\end{equation}
derived in the similar way as in Refs.~\onlinecite{PhysRevB.90.085132}
or \onlinecite{PhysRevB.78.054431}. Since $\gamma_{\text{vertical}}$ is
quantized into 0 or $\pi$ by the spatial inversion symmetry, and
$\langle S_z\rangle$ is fixed to 1/4 owing to the symmetry of the model,
$\gamma_{\text{diagonal}}$ is quantized into
$\pm\pi/2$. Figure~\ref{fig:type2}(b) indicates that the
topological
phase transition is captured as a jump in $\gamma$ since the bulk
symmetry is kept, but 0/$\pi$-quantization is broken because of the
symmetry breaking boundary.

The entanglement entropy is also plotted in
Fig.~\ref{fig:type2}. Different from the
case of 0/$\pi$ quantization, the entanglement entropy is finite in the
both of the positive and negative sides of the $\Delta=\pm 1$ limit. 
As in the case of the vertical edge, the entanglement entropy is 
obtained for $\Delta=\pm 1$ as $f(1/4)$. (See
Appendix~\ref{appendix}). Although the topological phase transition is
detected as a diverging behavior due to the bulk nature of the
entanglement entropy, degeneracy of the entanglement entropy
originating from the inversion symmetry does not take place for the
symmetry breaking boundary and we have a situation that the entanglement
entropy becomes smaller than $\log 2$, for instance,
$f(1/4)<f(1/2)=\log 2$. This behavior is confirmed in Fig.~\ref{fig:type2}. 

\begin{figure}[tbp]
 \begin{center}
  \includegraphics[scale=1.0]{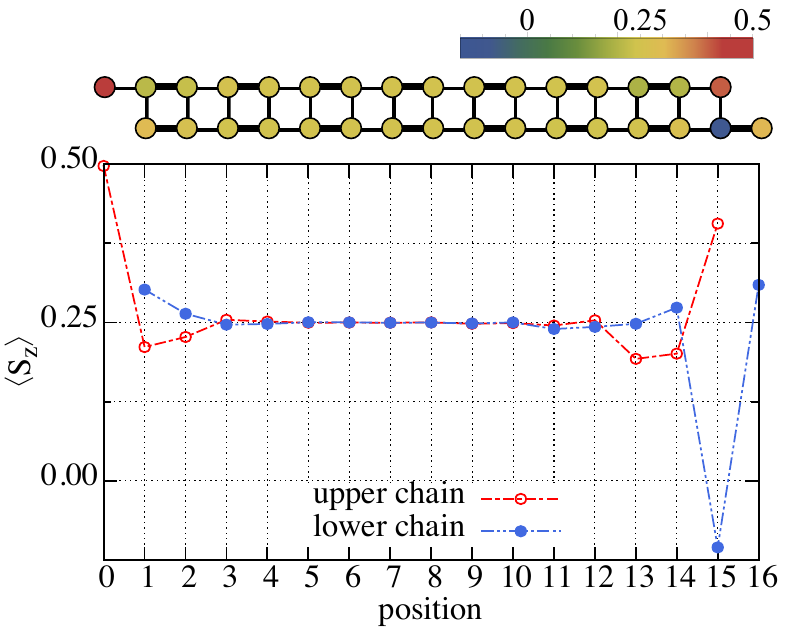}
  \caption{(Color online) The site resolved spin density in the
  1/2-plateau phase with
  the diagonal edge. The color map of the spin density is also
  shown. Calculation is performed on the system with 32 spins.}\label{fig:site_resolved}
 \end{center}
\end{figure}

Now we investigate the open ladder with the diagonal
edge to see the bulk--edge correspondence for the $\pm\pi/2$
quantization case. In Fig.~\ref{fig:site_resolved}, 
 the site resolved 
$\langle S_z\rangle$ for the ground state of $\langle S_z\rangle=1/4$
sector with $\Delta=0.5$ is plotted. There, the extra up spins are
accumulated at the left edge, while the extra down spins are accumulated
at the right edge. For negative $\Delta$, where the edge states are not
observed for the case with 0/$\pi$ quantization, there are still edge
states but roles of the right and the left edges are reversed. This
is consistent with the fact that the entanglement entropy goes to finite
values for both of $\Delta=\pm 1$. 
To summarize, the edge states for the
$\pm\pi/2$-quantization have features such that i) the up and down spins
are accumulated at the opposite ends, and ii) the edge states appear
both for positive and negative $\Delta$. These features give a physical
picture of the $\pm\pi/2$-quantization. Firstly, since $+\pi/2$
and $-\pi/2$ differ only in sign and have same magnitudes, we expect the
similar behavior of the edge states for positive and negative $\Delta$,
which explains the feature ii). This makes a clear contrast to the case of
0/$\pi$-quantization, where $0$ and $\pi$ are essentially different, and
leads to the absence and existence of the edge states. To understand the
feature i), analogy to the electron system is helpful. In free
electron systems, there is a direct relation between the
electronic polarization and the Berry phase\cite{PhysRevB.47.1651}. By
regarding up spins as electrons and down spins as holes, the state in
Fig.~\ref{fig:site_resolved} corresponds to the electronically polarized
state associated with the finite Berry phase. Note that $\gamma=\pi$
represents the situation where the mean position of the electrons is at
the middle point between two lattice points, which means that there is
no electronic polarization even though the Berry phase is finite. Thus,
the quantization into $\pm\pi/2$ instead of 0/$\pi$ is essential to
observe the feature i).

In order to observe a unique quantization of the Berry phase described
above, the ladder structure and the applied magnetic field are
essential. First
of all, the ladder structure allows us to consider the diagonal edge for
which the fractional quantization is achieved. Further,
Eq.~\eqref{gamma_trans} implies that we need to look at the plateau phase
with $\langle S_z\rangle=1/4$ to have the fractional quantization. Although
the plateau phase under consideration can be mapped to the zero
magnetization phase in the XXZ chain in the strong rung coupling
limit, the fractional quantization is not expected in the XXZ chain
since it does not admit the diagonal edge. 

\subsection{Ring exchange}
\begin{figure}[tbp]
 \begin{center}
  \includegraphics{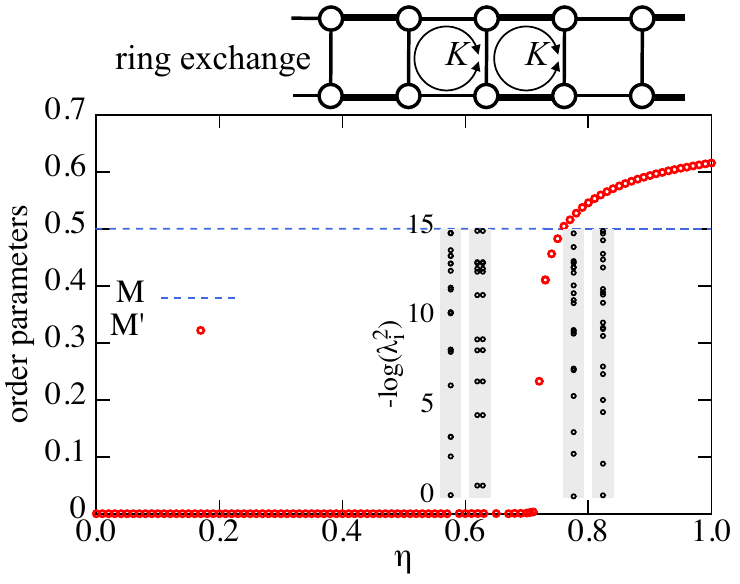}
  \caption{(Color online) Order parameters $M$ and $M'$ obtained with
  $B_z=2.0$ and $\chi=48$. Insets show
  the entanglement spectra for $\eta=0.6$ and $0.8$. For each value of $\eta$, the left column is for the edge on $J_1$ bond, and the right column is for the
  edge on $J_2$ bond.}\label{fig:ring}
 \end{center}
\end{figure}

So far, we have considered the 1/2-plateau phase induced by the
dimerization. For the uniform case without dimerization, a plateau
phase at the same magnetization $\langle S_z\rangle=1/4$ is induced by a
ring
exchange\cite{0953-8984-13-33-321,doi:10.1143/JPSJ.77.014709}, which 
is written as\cite{PhysRevB.60.48,0953-8984-13-33-321,PhysRevB.67.100409,PhysRevB.79.115107,PhysRevB.79.205107}
\begin{equation}
 \hat{H}_{\text{ring}}=K\sum_{i}(P_i+P^{-1}_i),
\end{equation}
where $P_i$ ($P_i^{-1}$) is an operator acting on the minimal four-site
plaquette that causes a clockwise (anticlockwise) shift of the spins
on that plaquette. That is, if we denote the states of
four spins on a plaquette as 
$
 \left|
 \begin{array}{cc}
  s_1&s_2\\
  s_3&s_4\\
 \end{array}
 \right\rangle
$, 
$P_i$ operates as 
\begin{equation}
 P_i
  \left|
 \begin{array}{cc}
  s_1&s_2\\
  s_3&s_4\\
 \end{array}
 \right\rangle
  =
 \left|
 \begin{array}{cc}
  s_2&s_4\\
  s_1&s_3\\
 \end{array}
 \right\rangle,\,
  P_i^{-1}
  \left|
 \begin{array}{cc}
  s_1&s_2\\
  s_3&s_4\\
 \end{array}
 \right\rangle
  =
 \left|
 \begin{array}{cc}
  s_3&s_1\\
  s_4&s_2\\
 \end{array}
 \right\rangle.
\end{equation}
Intuitive understanding of the ring exchange
is possible for the strong rung coupling limit. There, we have
noted that the model is effectively described by a XXZ chain model. The
ring exchange term modifies the anisotropy of the effective XXZ model,
i.e., the anisotropy is modified from the easy plane type to the easy
axis type (Ising type) for sufficiently large positive $K$. Then, the
antiferromagnetic order is developed in the effective model, and this
ordered phase corresponds to the 1/2-plateau phase in the original
model\cite{0953-8984-13-33-321}. Due to the antiferromagnetic nature of
the state,
the spatial inversion symmetry that is essential for the SPT phase is
broken. Then, a transition between
the SPT phase and the symmetry broken phase {\it within the 1/2-plateau
phase} is expected when the strength of the ring exchange
is suitably tuned. In fact, such a transition is observed in the present
model. In Fig.~\ref{fig:ring},
$M=\frac{1}{2L}\sum_{i}^{2L}\sum_{j=1,2}\langle S^z_{i,j}\rangle$, and
$M'=\frac{1}{2L}\sum_{i}^{2L}\sum_{j=1,2}(-1)^i\langle S^z_{i,j}\rangle$,
which captures the symmetry breaking, are plotted as
a function of $\eta$ introduced as $\Delta=0.2(1-\eta)$ and
$K=0.8\eta$. Note that the dimerization dominated plateau phase is
expected for $\eta=0$, while the ring exchange dominated plateau phase
is expected for $\eta=1$. The numerical result in Fig.~\ref{fig:ring}
confirms this idea. Namely, $M'$ gets finite at a certain value of
$\eta$ as $\eta$ increases, while $M$ is always 0.5. 

In order to clarify the difference between zero $M'$ phase
and finite $M'$ phase in terms of SPT, the entanglement spectra for
$\eta=0.6$ and $0.8$ are also shown in Fig.~\ref{fig:ring}. We did not use
the Berry phase since it requires a finite size system
for calculation, which means that it is difficult to treat the phase with
spontaneous symmetry breaking. Here, we only use
the vertical edges. Still, there are two choices of the vertical edges,
one breaks $J_1$ bonds and the other breaks $J_2$ bonds. Therefore, for
each value of $\eta$, two spectra are plotted. At $\eta=0.6$, we observe
double degeneracy of the spectrum for the edge on $J_2$ bonds.
This degeneracy stems from the nontrivial projective representation,
$\zeta=-1$, and signals a SPT phase. On the other hand, for $\eta=0.8$,
neither of the edges on $J_1$ and $J_2$ bonds leads to the degeneracy of
the spectra. These observation implies that the transition in
Fig.~\ref{fig:ring} is a typical
and concrete example for collapse of the SPT phase by a spontaneous
symmetry breaking. 

\subsection{Comparison with the 0-plateau phase}
Finally, we briefly discuss the 0-plateau phase near zero
magnetic field. In this case there is no sign of the phase transition in
both of the Berry phases and the entanglement entropy plotted as a
function of $\Delta$ (not shown). This is natural
because $\Delta=0$
state is not critical but described as a rung singlet state\cite{PhysRevB.67.100409,PhysRevB.74.155119} without the
Zeeman field. In contrast, the dimerized {\it chain} is critical when
$\Delta=0$ and $B_z=0$. However, actually, there is no jump at
$\Delta=0$ in the Berry phase as a function of $\Delta$, even in the
small rung coupling limit. This is because we apply gauge
twists on {\it all} bonds across the boundary. Then, even for $\Delta$
for which a topologically nontrivial phase is expected in the dimerized
{\it chain}, the Berry phase is 0 for the dimerized ladder because each
chain contributes $\pi$ to $\gamma$, which results in 
$\gamma=2\pi\equiv 0$ (mod $2\pi$). If we apply different
kinds of gauge twist, not limited to the one that is possible to be
reduced to a twisted boundary condition, it is possible to
detect dimer
structures\cite{PhysRevB.79.115107,PhysRevB.79.205107,PhysRevB.88.184418}. We
have also confirmed the absence of the topological phase protected by the
spatial inversion symmetry in 0-plateau phase by the MPS
representation. That is, in the 0-plateau phase, $\zeta$ is always 1
irrespective of the sign of $\Delta$.

\section{Summary}\label{sec:summary}
To summarize, it is established that there is a symmetry protected
topological phase in the 1/2-plateau phase in the dimerized spin-1/2
Heisenberg ladder. Even with the magnetic field, which is necessary to
access the 1/2-plateau phase and reduces the symmetry of the system, the
spatial inversion symmetry remains and protects the topological phase.
Namely, the inversion symmetry makes the Berry phase quantized into $0$
or $\pi$, and allow us to regard the Berry phase as a topological order
parameter. The entanglement entropy is also used to characterize the
topological phase. In order to complement the Berry phase based
arguments, the topological order parameter other than the Berry
phase is extracted from the MPS representation of the state. That
topological order parameter determines the degeneracy of the
entanglement spectrum, and gives lower bound of the entanglement
entropy. The importance of the symmetry is demonstrated by
introducing a symmetry breaking term and by spontaneous symmetry
breaking caused by the ring exchange. Because of the importance of the
inversion symmetry, the boundary that respects the inversion symmetry
is mainly used to define the Berry phase and the entanglement
entropy. However, our analysis on a specific
shape of boundary reveals that the symmetry breaking boundary can
lead to a new type of the bulk--edge correspondence. In specific, we
find a fractional quantization of the
Berry phase into $\pm\pi/2$ for such a symmetry breaking
boundary. Further it is shown that there are unique edge states, which
shows polarization, i.e., up and down spins are accumulated at the
opposite ends of the finite system, for the case with
$\pm\pi/2$-quantization. This finding implies new possible applications
of the Berry phase for exploring the topological properties of given
systems. For instance, the idea of the fractional quantization of the
Berry phase will have some applications also in two- or
three-dimensional systems in the magnetic fields. 
Other types of the fractional quantization can be possible for the
different plateau phases.

\begin{acknowledgments}
 This work is partly supported by Grants-in-Aid for Scientific Research,
 Nos. 26247064 and 25610101 from JSPS, and No. 25107005 from MEXT. 
 The authors thank the Supercomputer Center, the Institute for Solid
 State Physics, the University of Tokyo for the use of the facilities.
\end{acknowledgments}

\appendix
\section{Analytical calculation of the entanglement entropy for a
 decoupled four site cluster}\label{appendix}
\subsection{Hamiltonian and ground state}
For a four-site plaquette with $\langle S_z\rangle=1/4$, there are four
basis states written as
\begin{align}
 |\textrm{I}\rangle&=
  \left|
   \begin{array}{cc}
    \uparrow&\uparrow  \\
    \uparrow&\downarrow
   \end{array}
  \right\rangle,
 &|\textrm{II}\rangle&=
  \left|
   \begin{array}{cc}
    \uparrow&\uparrow  \\
    \downarrow&\uparrow
   \end{array}
  \right\rangle,\nonumber \\
 |\textrm{III}\rangle&=
  \left|
   \begin{array}{cc}
    \uparrow&\downarrow  \\
    \uparrow&\uparrow
   \end{array}
  \right\rangle,
 &|\textrm{IV}\rangle&=
  \left|
   \begin{array}{cc}
    \downarrow&\uparrow  \\
    \uparrow&\uparrow
   \end{array}
  \right\rangle.
\end{align}
The Hamiltonian of the plaquette acts on these basis states as
\begin{align}
 \hat{H}_p|I\rangle&=\frac{J_1}{2}|II\rangle+\frac{J_0}{2}|III\rangle,\\
 \hat{H}_p|II\rangle&=\frac{J_1}{2}|I\rangle+\frac{J_0}{2}|IV\rangle,\\
 \hat{H}_p|III\rangle&=\frac{J_1}{2}|IV\rangle+\frac{J_0}{2}|I\rangle,\\
 \hat{H}_p|IV\rangle&=\frac{J_1}{2}|III\rangle+\frac{J_0}{2}|II\rangle.
\end{align}
If $J_1>0$ and $J_0>0$, the ground state is obtained as
\begin{equation}
 |G\rangle=\frac{1}{2}
  \bigl(|I\rangle-|II\rangle-|III\rangle+|IV\rangle\bigr).
\end{equation}

\subsection{Entanglement entropy}
For a given bipartition, a state is expressed as
\begin{equation}
 |\psi\rangle=\sum_{ij}M_{ij}|\psi^<_i\rangle\otimes|\psi^>_j\rangle,
  \label{bipartition}
\end{equation}
where each of $|\psi^<_i\rangle$ and $|\psi^>_j\rangle$ is a state in
either part of the bipartitioned system. Using the singular value
decomposition, a matrix $\hat{M}$ can be always written as
\begin{equation}
 \hat{M}=\hat{U}\hat{\Lambda}\hat{V}^\dagger,
\end{equation}
where $\hat{\Lambda}$ is a diagonal matrix whose elements are
nonnegative, and $\hat{U}$ and $\hat{V}$ are unitary matrices. Then,
$|\psi\rangle$ is rewritten as
\begin{equation}
 |\psi\rangle=\sum_\alpha \lambda_\alpha 
  |\tilde{\psi}^<_\alpha\rangle\otimes
  |\tilde{\psi}^>_\alpha\rangle
\end{equation}
with 
\begin{equation}
 |\tilde{\psi}^<_\alpha\rangle=\sum_{i}U_{i\alpha}|\psi_i^<\rangle,\quad
 |\tilde{\psi}^>_\alpha\rangle=\sum_{j}(V^\dagger)_{\alpha j}|\psi_j^>\rangle,
\end{equation}
where $\lambda_\alpha$ denotes the diagonal elements of
$\hat{\Lambda}$. 

The entanglement entropy for this bipartition is evaluated as
\begin{equation}
 S=-\sum_\alpha \lambda_\alpha^2\log \lambda_\alpha^2.
\end{equation}

\subsection{Entanglement entropy for a plaquette}
First, we consider the vertical edge that breaks a four site plaquette
into two parts with two spins. Taking a set 
\begin{equation}
 \left|
\begin{array}{c}
 \uparrow\\
 \uparrow
\end{array}
 \right\rangle,\quad
 \left|
\begin{array}{c}
 \uparrow\\
 \downarrow
\end{array}
 \right\rangle,\quad
 \left|
\begin{array}{c}
 \downarrow\\
 \uparrow
\end{array}
 \right\rangle,\quad
 \left|
\begin{array}{c}
 \downarrow\\
 \downarrow
\end{array}
 \right\rangle,
\end{equation}
as $|\psi^<_i\rangle$ and $|\psi^>_i\rangle$, $|G\rangle$ can be written
in the form as Eq.~\eqref{bipartition} with $\hat{M}$ being 
\begin{equation}
 \hat{M}=
  \begin{pmatrix}
   0&\frac{1}{2}&-\frac{1}{2}&0\\
   -\frac{1}{2}&0 &0 & 0\\
   \frac{1}{2}&0&0&0\\
   0&0 &0 &0
  \end{pmatrix}.
\end{equation}
Then, it is straightforward to confirm that $\hat{U}$,
$\hat{V}^\dagger$, and $\hat{\Lambda}$ for this case become
\begin{equation}
 \hat{U}=
  \begin{pmatrix}
   -\frac{1}{\sqrt{2}}&-\frac{1}{2}&\frac{1}{2}&0\\
   \frac{1}{\sqrt{2}}&-\frac{1}{2}&\frac{1}{2}&0\\
   0&\frac{1}{\sqrt{2}} &\frac{1}{\sqrt{2}} & 0\\
   0&0&0&1
  \end{pmatrix},\quad
  \hat{V}^\dagger=
  \begin{pmatrix}
   \frac{1}{\sqrt{2}}&\frac{1}{\sqrt{2}}&0&0\\
   -\frac{1}{2}&\frac{1}{2}&\frac{1}{\sqrt{2}}&0\\
   \frac{1}{2}&-\frac{1}{2}&\frac{1}{\sqrt{2}}&0\\
   0&0&0&1
  \end{pmatrix},
\end{equation}
and 
\begin{equation}
   \hat{\Lambda}=
  \begin{pmatrix}
   \frac{1}{\sqrt{2}}&0&0&0\\
   0&\frac{1}{\sqrt{2}}&0&0\\
   0&0&0&0\\
   0&0&0&0
  \end{pmatrix}.
\end{equation}
As a result, we have
\begin{equation}
 S=-\frac{1}{2}\log\frac{1}{2}-\frac{1}{2}\log\frac{1}{2}
  =\log 2.
\end{equation}

Next, we consider the diagonal edge that breaks a four site plaquette
into two parts, one with a spin and another with three spins. Taking a
set 
\begin{equation}
 |\uparrow\rangle,\quad |\downarrow\rangle
\end{equation}
as $|\psi^<_i\rangle$ and a set
\begin{equation}
   \left|
   \begin{array}{cc}
    \uparrow&\uparrow  \\
    &\uparrow
   \end{array}
  \right\rangle,\quad
   \left|
   \begin{array}{cc}
    \uparrow&\uparrow  \\
    &\downarrow
   \end{array}
  \right\rangle,\quad
   \left|
   \begin{array}{cc}
    \uparrow&\downarrow  \\
    &\uparrow
   \end{array}
  \right\rangle,\quad
   \left|
   \begin{array}{cc}
    \downarrow&\uparrow  \\
    &\uparrow
   \end{array}
  \right\rangle,
\end{equation}
as $|\psi^>_i\rangle$, $|G\rangle$ is expressed as in the form of
Eq.~\eqref{bipartition} with
\begin{equation}
 \hat{M}=
  \begin{pmatrix}
   0&\frac{1}{2}&-\frac{1}{2}&\frac{1}{2}\\
   -\frac{1}{2}&0&0&0
  \end{pmatrix}.
\end{equation}
Now, it is easy to see that $\hat{U}$, $\hat{V}^\dagger$, and
$\hat{\Lambda}$ for this case are 
\begin{equation}
 \hat{U}=
  \begin{pmatrix}
   0&-1\\
   1&0
  \end{pmatrix},\quad
  \hat{V}^\dagger=
  \begin{pmatrix}
   1&0&0&0\\
   0&\frac{1}{\sqrt{3}}&0&\frac{1}{\sqrt{2}}\\
   0&-\frac{1}{\sqrt{3}} &\frac{1}{\sqrt{2}} &0\\ 
   0&\frac{1}{\sqrt{3}} &\frac{1}{\sqrt{2}} &-\frac{1}{\sqrt{2}}
  \end{pmatrix},
\end{equation}
and
\begin{equation}
  \hat{\Lambda}=
  \begin{pmatrix}
   \frac{1}{2}&0&0&0\\
   0&\frac{\sqrt{3}}{2}&0&0
  \end{pmatrix}. 
\end{equation}
Then, the entanglement entropy is obtained as
\begin{equation}
 S=-\frac{1}{4}\log\frac{1}{4}-\frac{3}{4}\log\frac{3}{4}.
\end{equation}

To summarize, we have $f(1/2)$ for the vertical edge and $f(1/4)$ for
the diagonal edge with $f(x)=-x\log x-(1-x)\log(1-x)$.

\end{document}